# Dry ice baths as liquid nitrogen substitutes for physics demonstrations

David Marasco, Foothill College, Los Altos Hills, CA
Paolo Marasco, San Jose State University, San Jose, CA

Liquid nitrogen (LN2) is a long-time favorite for physics demonstrations, with a large repertoire of crowd-pleasing experiments that are cornerstones in outreach efforts. While R1 universities usually have a ready LN2 supply for their Physics, Chemistry, and Biology departments, K-12 and two-year college teachers often have to go to specialty suppliers to obtain LN2, and also need access to expensive storage equipment. Dry ice (solid $CO_2$) is available at many super markets, and as such its suitability as a substitute for LN2 was explored, with the results discussed below. At 77K (-196°C) LN2 is considerably colder than dry ice at 195K (-78.5°C), however some demonstrations are still viable.

As many LN2 experiments involve submerging an object, that dry ice is a solid is an immediate concern. Phipps and Hume suggest the combination of dry ice and an organic solvent.[1] Although they warn against alcohols as these baths are highly viscous, isopropyl alcohol (rubbing alcohol) was used as like dry ice it is easily available. Experiments were performed with 91% isopropyl alcohol. Solutions were prepared in standard lab Pyrex beakers, and a selection of LN2 demonstrations commonly used in outreach[2-7] were tested. The demonstrations are categorized below:

**Shattering**
When brought to low temperature, many items are subject to shattering when struck. Common objects of interest include bananas, flowers, racquetballs, and pennies.

A banana was submerged in a one-liter dry ice bath for ten minutes. When it was removed and struck against a table it broke into several pieces (*fig 1*). As a medium-sized banana can have a large thermal mass, adding additional dry ice to the bath after the introduction of the banana is advisable. Flowers took about a minute, and shattered also (*fig 2*). However, despite a ten minute immersion, a racquet ball would not shatter, although it did bounce to a much reduced height when dropped. The bouncing behavior was carefully measured, and is described as a possible class activity later in this work.

Due to its geometry and heat capacity, a 2024 penny was placed directly on dry ice and allowed to cool, skipping the creation of a bath. The penny shattered when struck by a hammer (*fig 3*). As with LN2, while zinc (post-1981 pennies) become brittle, copper pennies do not shatter when struck. Ductile vs. brittle behavior in metals is related to the movement of dislocations in crystallographic planes.[8] Copper with a face-centered cubic (FCC) crystal structure has 12 different preferred planes where dislocations can occur. This large number makes copper ductile. Zinc has a hexagonal close-packed (HCP) crystal structure that features only 3 planes for dislocations which makes it more brittle, this behavior is more pronounced at lower temperature.

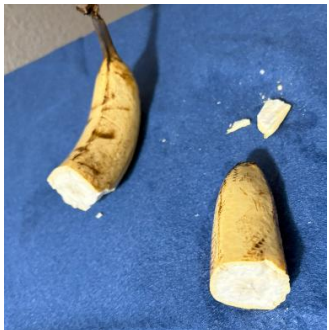
Figure 1: Banana after striking.

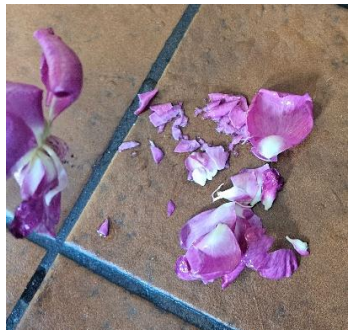
Figure 2: Shattered flower.

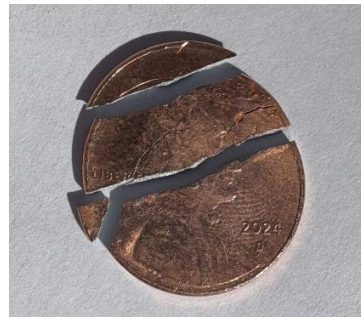
Figure 3: Zinc (post-1981) penny.

**Phase changes**

Another crowd favorite is immersing balloons in LN2 and watching them shrink. This demonstration will not work with a dry ice bath because most of the volume change occurs when the oxygen and nitrogen in the balloon liquify, which is not possible in this case as the temperature of dry ice is well above the phase change temperatures of those gases. If reduction in volume was due only to temperature, then applying the Ideal Gas Law would produce a volume at LN2 temperature that is 26% of the volume at room temperature (298K). This would correspond to a diameter that is 63% the size of room temperature diameter, contrary to the traditional results. The failure of the balloon demonstration was verified; as the gases in the balloon are not liquified, the balloon shrinks only via the Ideal Gas Law, which is a small effect (the theoretical diameter is 86% of room temperature). The classic demonstration where a small piece of dry ice is sealed into a balloon and inflates it as the dry ice warms and sublimates does show phase change with temperature, but is a less-exciting substitute for the balloons in LN2 experiment.

Another traditional demonstration is pouring LN2 into a tea kettle and listening to it whistle. This is due to the rapid evaporation and change of volume of the nitrogen as it encounters the relatively hot (room temperature) kettle. This demonstration was attempted in two stages. The first was to simply place dry ice into a kettle. This did not result in a fast enough production of $CO_2$ gas to create the whistling effect. However, when warm tap water was added to cover the dry ice, the kettle would whistle (*fig 4*).

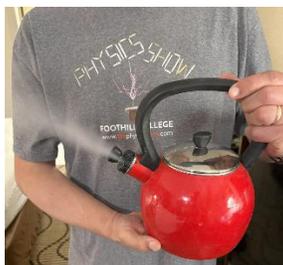
Figure 4: Whistling tea kettle.

Using LN2 to create instant ice cream from raw ingredients is a classic outreach activity. Dry ice can also serve in this role. While Coppola and coworkers[7] determined that a wide range of recipes are successful, their basic vanilla was selected due to the small number of ingredients:
- 4 cups milk
- 4 cups heavy cream
- 1⅓ cups granulated sugar (original recipe calls for superfine)
- 4 teaspoons vanilla

In a large metal bowl the sugar was blended into the liquids until it dissolved, and then the vanilla was added. The dry ice was pulverized inside its bag, and very small fragments were stirred into the mixture until the ice cream set (*fig 5*). It is important to reduce the dry ice to a kosher-salt grain size to avoid the possibility of any dry ice persisting under a coating of frozen dairy, lurking as a danger to eaters. During the initial stages, there was a lot of foam, so the larger the bowl, the better. The creation of the ice cream required roughly 20 minutes of manual stirring and between 3 and 4 pounds of dry ice. It is best if two people are involved in the production of the ice cream, one to stir, and the other to add dry ice, and also so they can switch roles as the stirring is tedious. In order to avoid overpressure situations, do not store leftover ice cream in sealed containers.

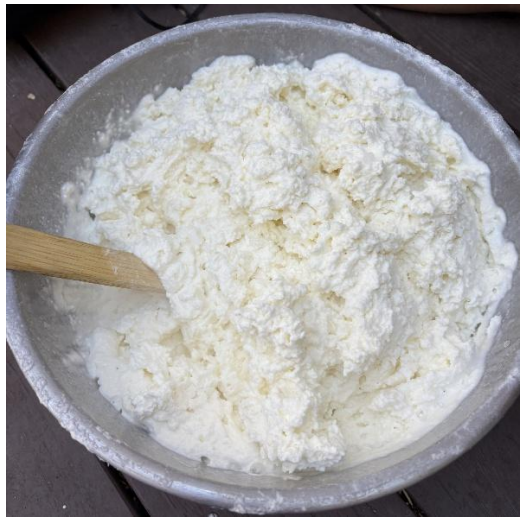

*Figure 5: Dry ice ice cream.*

**LED color**

When submerged in LN2 many LEDs undergo color shifts, some subtle, some dramatic. With LN2 the effect varies between different LEDs, depending on the details of the band gap in the semiconductors. The physics of this phenomena are covered briefly in the work of Planinšič and Etkina[4] and more deeply in Lisensky and coworkers.[9] A range of LEDs were tested, and while for many there was no detectable shift, there were some where there was a visible change in hue (*fig 6*), though none as notable as with LN2. For best results this is done in the dark, with a white background, with yellow LEDs.[10] If instructors have access to spectrometers, the change in wavelength can be directly measured (*fig 7*). With a good assortment of LEDs, dry ice is a tolerable stand-in for LN2.

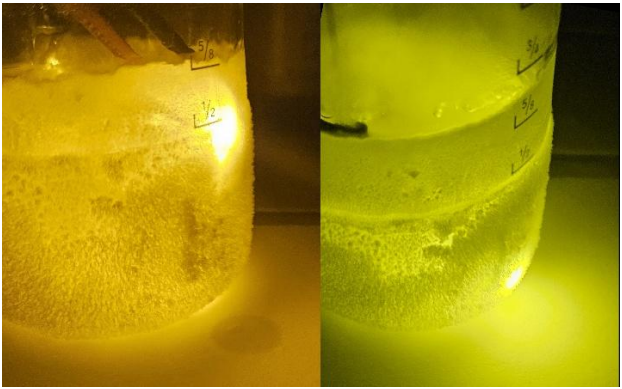
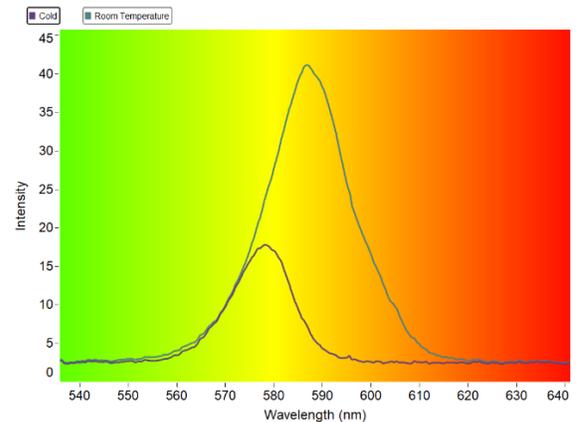

Figure 6: LED shifts color as it cools. LED introduced to the bath (left) and after reaching equilibrium (right).

Figure 7: Spectrometer data for a yellow LED at room-temperature (peak 586.7nm) and submerged in a dry ice bath (peak 578nm).

**Leidenfrost effect**
There are multiple implementations of the Leidenfrost Effect using LN2. People either hold a small amount of LN2 in their mouths or splash some on their skin. These are ideas highly not recommended for use with dry ice baths as experimenters should avoid isopropyl alcohol in their mouths, and the skin demonstration relies upon fast evaporation upon contact, which will not happen with the alcohol. The viscosity of the alcohol at low temperature can also be problematic.

A common LN2 experiment is cooling a piece of graham cracker, holding it in the mouth, and blowing "smoke" out of the mouth and nostrils. Due to the properties of isopropyl alcohol, the authors did not wish to cool the cracker by soaking in a bath. Making a reverse ice cream sandwich of a cracker between two pieces of dry ice did not sufficiently cool the cracker. Pieces of graham cracker were left in a bag of dry ice in a freezer overnight,[11] but only produced a small effect, not something noticeable at the distance of an audience. Dry ice is not a substitute for Leidenfrost effect demonstrations.

**Coefficient of restitution for a cooled racquetball**
While a racquetball will not shatter at dry ice temperatures, it does lose much of its bounce. Although the main focus of this work concerns physics demos for outreach, the reduced elasticity of the racquetball does open the door to a classroom or lab activity, the physics of a bouncing ball and its application to instruction has seen extensive research.[12-14] Slow-motion videos of a bouncing room-temperature racquetball and a racquet ball that had been cooled for 30 seconds in a dry ice bath were recorded. The cooled ball is not at dry ice temperature, but is cold enough to see an effect, yet not so cold as to make experimental data hard to analyze. PASCO's Capstone software[15] was used for video analysis, capturing the height of the ball for a series of bounces (*fig 8*). The data were exported to a spreadsheet and corrections were made for motion in the direction of the line of sight. With these data the coefficient of restitution for the ball can be measured. The coefficient of restitution is an important figure of merit from the study of collisions, and is defined as:

$$\varepsilon = v_f / v_i \qquad (1)$$

Where $\varepsilon$ is the coefficient of restitution and $v_f$ and $v_i$ are the relative velocities of the objects exiting and entering a collision. In the simplified context of dropping a ball, the two velocities refer to the velocities of the ball as the Earth is assumed motionless. As $\varepsilon^2$ is the ratio of kinetic energy after the collision to kinetic energy before, conservation of energy implies that $\varepsilon$ can be measured by taking the square root of the ratio of the height of a bounce to the height of the previous bounce. For the room temperature ball the coefficient of restitution was 0.91+/-0.02 whereas for the cooled ball it was 0.749+/-0.007, thus the ball will lose roughly 19% of its energy on each bounce when warm, but 44% when cold. Video captures of the bounces and the resulting spreadsheets for these experiments are provided as a resource for instructors who wish to do this exercise in class but do not have the resources to take the data themselves.[16]

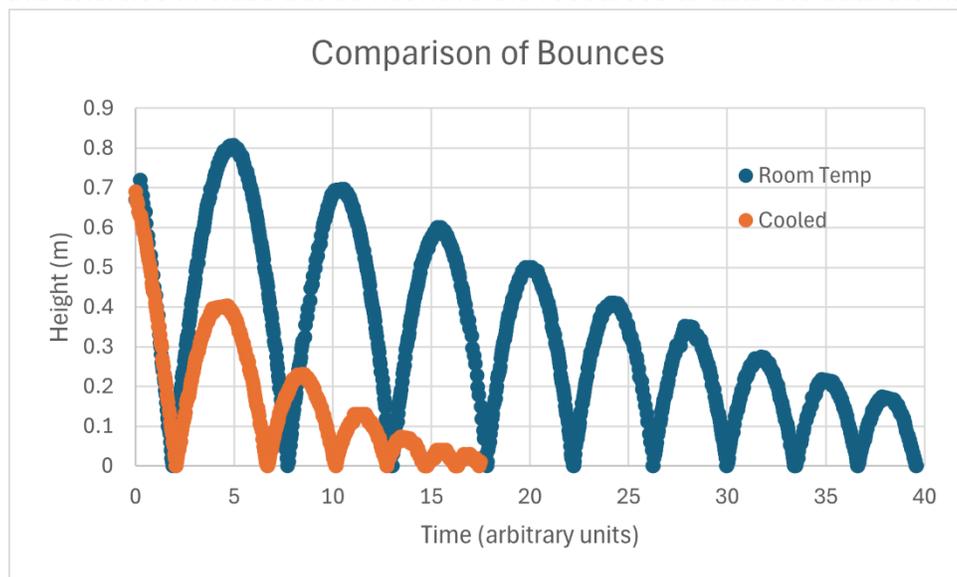

*Figure 8: Height as a function of time for the room temperature bouncing racquetball and the cooled bouncing racquetball.*

## Conclusion
While dry ice cannot act as a complete replacement for LN2, there are a range of physics demonstrations where it can act as a stand-in. Given the low cost and ready availability of the materials, this technique is a resource for those who do not have access to LN2.

## Acknowledgements
The authors wish to thank the reviewers, whose feedback greatly strengthened this work. In addition, we wish to recognize the physics outreach community which has finely honed many of the demonstrations examined.

**References**

1. Alan M. Phipps and David N. Hume, General purpose low temperature dry ice baths, J. Chem. Educ. 1968, 45, 10, 664. https://doi.org/10.1021/ed045p664

2. J. C. Sprott, *Physics Demonstrations: A Sourcebook for Teachers of Physics,* 1st ed. (U. Wisconsin Press, Madison, 2006). P. 88-92

3. R. G. Hunt, G. L. Salinger; TPT NOTES: Qualitative Demonstrations and Experiments Using Liquid Nitrogen. *Phys. Teach.* 1 May 1969; 7 (5): 289–290. https://doi.org/10.1119/1.2351369

4. Gorazd Planinšič, Eugenia Etkina; Light-Emitting Diodes: Learning New Physics. *Phys. Teach.* 1 April 2015; 53 (4): 210–216. https://doi.org/10.1119/1.4914558

5. J. S. Miller, *Science Demonstrations: Teacher's Manual,* 1st ed. (Western Instructional Television Inc., Los Angeles, 1979). P. 52

6. Jearl Walker. "Boiling and the Leidenfrost effect." *Fundamentals of physics* (2010): E10-1.

7. B.P.Coppola, J.W. Hovick, and D.S. Daniels; I Scream, You Scream…: A New Twist on the Liquid Nitrogen Demonstrations. J. Chem. Educ. 1 December 1994; 71 (12), 1080. https://pubs.acs.org/doi/abs/10.1021/ed071p1080

8. William D. Callister, Jr. and David G. Rethwisch, *Materials Science and Engineering: An Introduction,* 10th ed (Wiley, New Jersey, 2018), p. 185-6

9. George C. Lisensky, R. Lee Penn, Margret J. Geselbracht, and Arthur B. Ellis; Periodic properties in a family of common semiconductors: Experiments with light emitting diodes. *J. Chem. Educ.* 1 February 1992; 69 (2): 151–156. https://doi.org/10.1021/ed069p151

10. The LED used was the yellow option from the BOJACK 200 piece LED kit (BJ-L10-200). Our experience is that vendors may change their suppliers, so a part number is not a guarantee of getting the same LED over time.

11. While some dry ice may persist overnight in a normal freezer, a drawback is that it cannot be stored for any appreciable amount of time at that temperature.

12. Rod Cross; The coefficient of restitution for collisions of happy balls, unhappy balls, and tennis balls. Am. J. Phys. 1 November 2000; 68 (11): 1025–1031. https://doi.org/10.1119/1.1285945

13. Kelley D. Sullivan; What's in a Name: Why Do We Call a Bouncy Ball Bouncy?. Phys. Teach. 1 April 2019; 57 (4): 229–231. https://doi.org/10.1119/1.5095376

14. David Marasco; Teaching an Old Ball New Tricks: Another Look at Energetics, Motion Detectors, and a Bouncing Rubber Ball. Phys. Teach. 1 January 2020; 58 (1): 62–63. https://doi.org/10.1119/1.5141977

15. https://www.pasco.com/products/software/capstone

16. https://huggingface.co/datasets/marascodavid/DryIceRubberBall/tree/main Note that the videos do not have sound.